\documentclass[%
 reprint,
 amsmath,amssymb,
 aps,
 prapplied,
superscriptaddress]{revtex4-2}
\usepackage{lipsum}
\usepackage{graphicx}
\usepackage{dcolumn}
\usepackage{bm}
\usepackage{xcolor}
\usepackage{newtxtext,newtxmath}
\usepackage{siunitx}
\definecolor{changecolor}{RGB}{0, 0, 0} 
\begin{document}

\preprint{APS/123-QED}

\title{\textcolor{changecolor}{Bridging gaps in Rydberg RF receivers using modulation transfer bandwidth enhancement}}

\author{Mickael Branco}
\affiliation{
Université Paris-Saclay, CNRS, ENS Paris-Saclay,
CentraleSupélec, LuMIn, 91400 Orsay, France}
\author{K V Adwaith}
\affiliation{
Université Paris-Saclay, CNRS, ENS Paris-Saclay,
CentraleSupélec, LuMIn, 91400 Orsay, France}
\author{Gabriel Boccara}
\affiliation{
Université Paris-Saclay, CNRS, ENS Paris-Saclay,
CentraleSupélec, LuMIn, 91400 Orsay, France}
\author{Duc-Anh Trinh}
\affiliation{
Université Paris-Saclay, CNRS, ENS Paris-Saclay,
CentraleSupélec, LuMIn, 91400 Orsay, France}

\author{Sacha Welinski}
\affiliation{
 Thales Research and Technology, 91120 Palaiseau, France
}
\author{Perrine Berger}
\affiliation{
 Thales Research and Technology, 91120 Palaiseau, France
}

\author{Fabienne Goldfarb}
\affiliation{
Université Paris-Saclay, CNRS, ENS Paris-Saclay,
CentraleSupélec, LuMIn, 91400 Orsay, France}
\author{Fabien Bretenaker}
\affiliation{
Université Paris-Saclay, CNRS, ENS Paris-Saclay,
CentraleSupélec, LuMIn, 91400 Orsay, France}
\date{\today}
\begin{abstract}
\textcolor{changecolor}{We optimize theoretically and experimentally the performances of the} recently demonstrated modulation transfer protocol [D.-A. Trinh, K. V. Adwaith, M. Branco, A. Rouxel, S. Welinski,
P. Berger, F. Goldfarb, and F. Bretenaker, Applied Physics Letters \textbf{125},
154001 (2024)] aiming at extending the bandwidth of quantum RF receivers based on hot Rydberg atoms. \textcolor{changecolor}{This optimization relies on tuning the parameters of the phase modulation of the coupling beam, which is converted by the nonlinear response of the atoms into an amplitude modulation of the probe beam.}
We develop a theoretical model to optimize both the modulation frequency and the modulation amplitude of the coupling beam, thereby maximizing the atomic response. Once optimized, the sensitivity to detuned RF fields of this modulation transfer protocol is \textcolor{changecolor}{measured and} compared with that of the conventional protocol. This comparison shows that the new protocol \textcolor{changecolor}{permits a strong increase in the detection bandwidth. Indeed, it} outperforms the usual one as soon as the RF signal to be measured is detuned by more than a few MHz. \textcolor{changecolor}{We illustrate the capability of this modulation transfer protocol to enhance the detection bandwidth by showing experimentally how it permits to bridge the gap between two  Rydberg transitions separated by 166\,MHz.} In all cases, the experimental results are in good agreement with the simulations.
\end{abstract}

\maketitle

\section{\label{sec:level1}Introduction}

A large fraction of quantum sensors are based on atomic resonances probed by electromagnetic radiation. In such cases, the sensor sensitivity typically depends on i) the strength of the probed resonance and ii) the linewidth of the resonance \cite{budkerOpticalMagnetometry2007}. According to these two criteria, Rydberg atoms, i.e. atomic levels corresponding to large values of the principal quantum number $n$, are particularly appealing because i) the electric dipole moment between two neighboring Rydberg levels scales like $n^2$ and ii) the lifetime of Rydberg levels can be as long as milliseconds, making resonances between them extremely narrow. 

Since the transitions between Rydberg levels exhibit frequencies ranging from the MHz to the THz domain, these considerations have led to the development of very promising RF receivers based on such Rydberg levels. These RF receivers rely on a three-level Electromagnetically Induced Transparency (EIT) \cite{finkelsteinPracticalGuideElectromagnetically2023} in a thermal vapor of alkali atoms, where the RF field induces an Autler-Townes (AT) splitting. This technique was initially introduced using Rubidium-87 ($\mathrm{^{87} Rb}$) \cite{sedlacekMicrowaveElectrometryRydberg2012} and has since been expanded to include Rubidium-85 ($\mathrm{^{85} Rb}$) \cite{hollowayBroadbandRydbergAtomBased2014} and Cesium \cite{fanDispersiveRadioFrequency2016}.

While the primary focus of such sensors is electrometry through the measurement of RF field amplitude, some studies have developed alternative methods using the same initial apparatus. These methods aim to measure the phase \cite{simonsRydbergAtombasedMixer2019,andersonOpticalRadioFrequencyPhase2022,jingAtomicSuperheterodyneReceiver2020}, the polarization \cite{sedlacekAtomBasedVectorMicrowave2013, wangPreciseMeasurementMicrowave2023}, or even the angle of arrival of the RF field \cite{robinsonDeterminingAngleofarrivalRadiofrequency2021,yanThreedimensionalLocationSystem2023}.

This all-optical detection method allows for detecting signals ranging from the sub-Megahertz \cite{jauVaporCellBasedAtomicElectrometry2020, liuHighlySensitiveMeasurement2022,limKilohertzrangeElectricField2023, brownVeryhighUltrahighfrequencyElectricfield2023} range to Terahertz frequencies \cite{chenTerahertzElectrometryInfrared2022}, exploiting the large number of allowed atomic transitions between Rydberg states in alkali atoms. These transitions exhibit large electric dipole moments, which make Rydberg RF receivers highly sensitive to resonant microwave fields.

However, this sensitivity decreases very rapidly when the RF signal is detuned from the narrow resonance. Indeed, while the dependence of the energies of the dressed states is linear with the RF field amplitude when it is exactly at resonance (Autler-Townes doublet), this dependence becomes quadratic (light shift) far from resonance. 
Additionally, there are notable gaps in the Gigahertz range where transitions are separated by several megahertz. This limitation restricts the ability to achieve continuous frequency detection.

To address this limitation, some research groups have proposed to improve the RF bandwidth by inducing an AC Stark shift using a high-intensity \cite{andersonContinuousfrequencyMeasurementsHighintensity2017} detuned RF field. This approach can create a large bandwidth of several Gigahertz (ranging from 2 to 5 GHz), but results in a reduced sensitivity, which is not ideal for detecting weak RF fields. Another approach consists in using an RF local oscillator \cite{huContinuouslyTunableRadio2022,yaoSensitivityEnhancementFardetuned2022,liRydbergAtombasedAM2022} to enhance the sensitivity and enable detection of weak RF fields. It has also been demonstrated that exciting a second  Rydberg transition permits to tune the sensor resonance frequency \cite{simonsContinuousRadiofrequencyElectricfield2021}. However, both of these techniques require a second RF field, which undermines the goal of developing an all-optical detection system. Alternatively, DC Stark shifts \cite{ouyangContinuousBroadbandMicrowave2023} can be used, but they necessitate electrodes inside the vapor cell, which also contradicts the aim of creating an all-dielectric detection system.

In order to overcome these limitations, an all-optical method has been demonstrated to increase the RF bandwidth \textcolor{changecolor}{by using a modulation transfer similar to four-wave mixing }\cite{heNoiseSpectroscopyRydberg2021,trinhModulationTransferProtocol2024}. It is based on the use of a phase modulation of the coupling laser, which is transferred to the probe laser through the nonlinearity of the atomic medium. A new signal is then obtained by detecting the intensity modulation of the probe field, which has been shown to allow a significant improvement of the RF detection bandwidth \cite{trinhModulationTransferProtocol2024}.

\textcolor{changecolor}{The aim of this article is thus to improve this modulation transfer protocol (MTP) to expand the RF bandwidth as much as possible compared to the conventional protocol (CP), thereby closing the gap between two closely spaced RF transitions.} This is performed by optimizing the modulation parameters in a three-level ladder EIT scheme in $\mathrm{^{85} Rb}$. 
\textcolor{changecolor}{The first step consists in developing and using a reliable theoretical model. This is performed in Section \ref{sec:theory}, where we simulate the response of the system along the CP and the MTP.} This model is subsequently used to optimize the modulation parameters. The results from these simulations are then compared with the experimental measurements in Section \ref{sec:optimization}. \textcolor{changecolor}{Then, Section \ref{section:bandwidth} provides a detailed comparison between the performances of the two protocols. The signal and the noise are carefully measured and analyzed to determine the conditions in which the MTP outperforms the usual approach. A complete comparison of the bandwidths and sensitivities is provided, and substantiated with simulations. Finally, the interest of the new approach is illustrated by showing how it permits maintaining a good sensitivity while the RF frequency is tuned between two adjacent Rydberg transitions, whereas the conventional approach leads to a complete loss of sensitivity.}

\section{Theory and simulations}\label{sec:theory}

As stated in the introduction, this section is dedicated to the theory and simulation of the Rydberg sensor along the two protocols. In the conventional protocol (CP, Section II.A), variations in DC transmission of the probe field are used to measure the RF field. In the modulation transfer protocol \cite{trinhModulationTransferProtocol2024}, the coupling beam is phase modulated leading to a modulation of the probe beam intensity (Section II.B). \textcolor{changecolor}{The simulations detailed in this section will be used in Section III.B to optimize the modulation parameters.}
\subsection{Conventional Protocol}

\begin{figure}[tbp]
\includegraphics[width = \columnwidth]{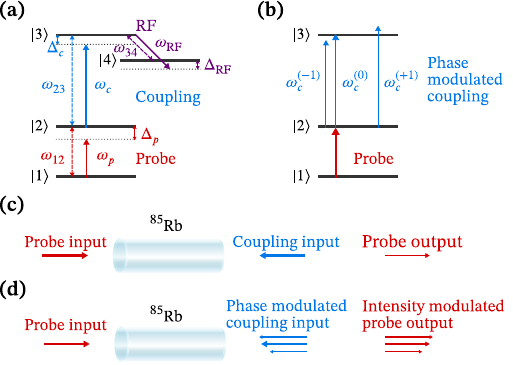}
\caption{\label{fig:principle} (a) Excitation scheme used for the conventional Rydberg RF receiver. (b) Excitation scheme used for the MTP Rydberg RF receiver, where the coupling field is phase modulated. (c) CP featuring single-frequency coupling and output probe fields. (d) MTP, including sidebands for both the coupling and output probe fields.}
\end{figure}

To model the interaction between the atoms and the electromagnetic fields, we consider a four-level atom as schematized in Fig.\,\ref{fig:principle}(a). The atom is initially in the ground state  $\left|1\right>$  and is coupled to the first excited state $\left|2\right>$ by a probe laser with frequency $\omega_{\mathrm{p}}$. Additionally, a coupling laser with frequency  $\omega_{\mathrm{c}}$ connects this latter level to the Rydberg state $\left|3\right>$. The electric fields associated with the two counter-propagating lasers (see Fig.\,\ref{fig:principle}(b)) read $E_{\mathrm{p}}(x, t) = \mathcal{E}_{\mathrm{p}}\,\mathrm{e}^{-\mathrm{i}\omega_{\mathrm{p}} t+\mathrm{i} k_{\mathrm{p}} x} + c.c.$ and $E_{\mathrm{c}}(x, t) = \mathcal{E}_{\mathrm{c}}\,\mathrm{e}^{-\mathrm{i}\omega_{\mathrm{c}} t-\mathrm{i} k_{\mathrm{c}} x} + c.c.$, respectively. Moreover, the radio frequency (RF) field to be detected couples the atom to a two Rydberg state $\left|3\right>$ and $\left|4\right>$ with a frequency $\omega_{\mathrm{RF}}$. Its electric field is given by: $E_{\mathrm{RF}}(z, t) = \mathcal{E}_{\mathrm{RF}}\,\mathrm{e}^{-\mathrm{i}\omega_{\mathrm{RF}} t -\mathrm{i}k_\mathrm{RF}z}+ c.c$. 

The Hamiltonian $H$ that drives the system consists of an atomic part $H_\mathrm{A}$ and an interaction term defined as $H_\mathrm{I} ~=~ - \boldsymbol{\hat\wp}  \cdot \bm{E}_{\mathrm{total}}$, where $\boldsymbol{\hat\wp}$ is the dipole operator.
The different Rabi frequencies are $\Omega_{\mathrm{p}} = 2\wp_{21}\mathcal{E}_{\mathrm{p}}/\hbar$ for the probe field, $\Omega_{\mathrm{c}}~=~2\wp_{32}\mathcal{E}_{\mathrm{c}}/\hbar$ for the coupling field, and $\Omega_\mathrm{RF} = 2\wp_{34}\mathcal{E}_\mathrm{RF}/\hbar$ for the RF wave. We introduce the detunings  $\Delta_{21} = \omega_{\mathrm{p}}-\omega_{21}$,  $\Delta_{31} =( \omega_{\mathrm{p}} + \omega_{\mathrm{c}}) -\omega_{31}$, and $\Delta_{41} = (\omega_{\mathrm{p}}+\omega_{\mathrm{c}} - \omega_\mathrm{RF}) -\omega_{41}$ where  $\omega_{ij}$is the transition frequency between levels $\left|i\right>$ and $\left|j\right>$. In the frame rotating with the fields, the Hamiltonian governing the system evolution then becomes:
\begin{equation}
     \tilde{H} = -\frac{\hbar}{2}\begin{pmatrix} 0 & \Omega_{\mathrm{p}}^* & 0 & 0\\
    \Omega_{\mathrm{p}} & 2\Delta_{21} & \Omega_{\mathrm{c}}^* & 0\\
    0 & \Omega_{\mathrm{c}} & 2\Delta_{31} & \Omega_\mathrm{RF}^* \\
    0 & 0 & \Omega_\mathrm{RF} & 2\Delta_{41}\end{pmatrix} \ .
\end{equation}
The density operator describing the state of the atoms evolves according to the von Neumann equation
\begin{equation}
    \frac{\mathrm{d}\rho}{\mathrm{d}t} = - \frac{i}{\hbar} [\tilde{H}, \rho ] + \mathcal{L}\ ,\label{Eq01} 
\end{equation}
where $\mathcal{L}$ models the relaxation mechanisms. The decay rate for the population $\rho_{ii}$ of level $|i\rangle$ is noted $\Gamma_i$ while $\gamma_{ij} = \frac{1}{2}(\Gamma_i + \Gamma_j)$ is the decay rate for the coherences $\rho_{ij}$ between levels $|i\rangle$ and  $|j\rangle$. The transit rate of the atoms through the laser beams is noted $\gamma_{\mathrm{t}}$, and adds to the decay rates of all the elements of the density matrix. Finally, The rate of atoms in the ground state that enter the interaction region is noted $\gamma_{\mathrm{in}}$.
By casting the density matrix $\rho$ as a 16-component column vector, Eq.\,(\ref{Eq01}) can be re-written in the following form:
\begin{equation}\label{eq:1}
    \frac{\mathrm{d}\rho}{\mathrm{d}t} = M\rho - R\rho + N\ ,
\end{equation}
where $M$ is a $16\times16$ matrix containing the elements coming from the Hamiltonian, $R$ is also a $16 \times16$ matrix containing the relaxation terms, and $N$ is a 16-element column vector in which the first element is $\gamma_\mathrm{in}$ and the other elements are zero. In the following, we look for the steady-state solutions of Eq.\,(\ref{eq:1}).

The evolution of the probe field during its propagation in the atomic medium is obtained from Maxwell equations. The polarization of the atoms at the probe frequency is $P_{\mathrm{p}}(\mathbf{r},t)~=~ \mathcal{P}_{\mathrm{p}}\,{\mathrm{e}}^{-{\mathrm{i}}(\omega_{\mathrm{p}}t-k_{\mathrm{p}}x)}+c.c.$ with a complex amplitude $\mathcal{P}_{\mathrm{p}}(x,t) ~=~  \mathcal{N}_0\,\wp_{12}\,\rho_{21}(x,t)$, where $\mathcal{N}_0$ is the atomic density. In order to take into account the large optical thickness of the atomic cell at the probe wavelength, we divide its thickness into a large number of small slices of thickness $dz$, each of them corresponding to a weak linear absorption smaller than 1\%. Then, for each of these slices, the evolution of the probe field amplitude reads: 
\begin{equation}\label{eq:2}
    \mathcal{E}_{\mathrm{p}}(x+dx) = \mathcal{E}_{\mathrm{p}}(x) + \frac{\mathrm{i}\omega_{\mathrm{p}}}{2\varepsilon_0c}\mathcal{N}_0\wp_{12}dx\int_{-\infty}^{+\infty}\rho_{21}(x,v)P(v)dv\ ,
\end{equation}
where $c$ is the velocity of light and $P(v)$ is the Maxwell distribution of the longitudinal components $v$ of the atomic velocities. In the integral of Eq.\,(\ref{eq:2}), the optical coherence $\rho_{21}(x,v)$ depends on the atomic velocity $v$ through the detunings of the probe and the coupling beams, which are affected by the Doppler shifts $-k_{\mathrm{p}}v$ and $k_{\mathrm{c}}v$, respectively.

In the experiments described below, the atomic medium is optically thick, and we divide its length $L$ into typically 100  slices of equal thickness $dx$, each absorbing less than $1\%$ of the probe power. The output of each serves as the input for the next one. We then obtain the probe transmission as $|\mathcal{E}_{\mathrm{p}}(L)|^2\big/|\mathcal{E}_{\mathrm{p}}(0)|^2$.

In the following, we focus on the following levels of $^{85}$Rb: level $\left|1\right>$ is $5^2S_{1/2}(F= 3)$, level $\left|2\right>$ is $ 5^2P_{3/2}(F = 4)$, level $\left|3\right>$ is $50^2D_{5/2}(F=3,4,5)$, and level $\left|4\right>$ corresponds to $51^2P_{3/2}(F=2,3,4)$. The values of parameters $\omega_{ij}$, $\wp_{ij}$, and $\Gamma_{i}$ are obtained from the Python package called Alkali Rydberg Calculator (ARC) \cite{sibalicARCOpensourceLibrary2017}. The values of the elements of the electric dipole matrix are obtained by averaging over the transitions between the different sublevels that constitute each transition. The values of these parameters used in the simulations are compiled in Table \ref{tab01}. Moreover, the small signal transmission of the cell at resonance is equal to 0.34, and the atom transit rate in the beam is  $\gamma_{\mathrm{t}}/2\pi=650\,\mathrm{kHz}$. All these values are deduced from the parameters of the experiment described below (see Section \ref{sec:optimization}.A).

\begin{table}
\caption{Parameters used in the simulations (see Fig.\,\ref{fig:Figure2}) which correspond to the experimental parameters (see Fig.\,\ref{fig:exp_MTP}) introduced in Section \ref{sec:optimization}.A}
\label{tab01}
\begin{ruledtabular}
\begin{tabular}{cccccc}
 Dipole &    & Population &   &  Rabi &    \\
 \textcolor{changecolor}{Moment} &  ($ea_0$) & \textcolor{changecolor}{Decay} &  (MHz) & Frequency &  (MHz) \\
\hline
$\wp_\mathrm{12}$ & $1.96$ & $\Gamma_2/2\pi$ & $6.050$ & $\Omega_\mathrm{p}/2\pi$ & $1.32$ \\
$\wp_\mathrm{23}$ & $0.01$ & $\Gamma_3/2\pi$ & $0.002$ &  $\Omega_\mathrm{c}/2\pi$ & $2.38$\\
$\wp_\mathrm{34}$ & $2272.4$ & $\Gamma_4/2\pi$ & $0.002$& $\Omega_\mathrm{RF}/2\pi$ & Variable\\
\end{tabular}
\end{ruledtabular}
\end{table}

\subsection{Modulation Transfer Protocol}
In the modulation transfer protocol, the coupling laser is phase modulated at frequency $\omega_\mathrm{mod}$. This creates two sidebands with opposite signs as shown in Fig.\,\ref{fig:principle}(b).  
The coupling field then reads $E_{\mathrm{c}}(x, t)= E_{\mathrm{c}}^{(-1)}(x, t) + E_{\mathrm{c}}^{(0)}(x, t) + E_{\mathrm{c}}^{(+1)}(x, t)$  where the carrier and sideband field components respectively read:
\begin{equation}
\left\lbrace
\begin{aligned}
    E_{\mathrm{c}}^{(0)}(x, t) &= \mathcal{E}_{\mathrm{c}}^{(0)} \mathrm{e}^{-\mathrm{i}(\omega_{\mathrm{c}}t+k_{\mathrm{c}}x)}+ \text{c.c.}, \\
    E_{\mathrm{c}}^{(-1)}(x, t) &= \mathcal{E}_{\mathrm{c}}^{(-1)}\mathrm{e}^{\mathrm{i}\omega_{\mathrm{mod}}t}\mathrm{e}^{-\mathrm{i}(\omega_{\mathrm{c}}t+k_{\mathrm{c}}x)} + \text{c.c.}, \\
    E_{\mathrm{c}}^{(+1)}(x, t) &= \mathcal{E}_{\mathrm{c}}^{(+1)}\mathrm{e}^{-\mathrm{i}\omega_{\mathrm{mod}}t}\mathrm{e}^{-\mathrm{i}(\omega_{\mathrm{c}}t+k_{\mathrm{c}}x)} + \text{c.c.},
\end{aligned}
\right.
\end{equation}
where $\mathcal{E}_{\mathrm{c}}^{(+1)}$ and $\mathcal{E}_{\mathrm{c}}^{(-1)}$ are the complex amplitudes of the sidebands. Due to the presence of these sidebands, Eq. (\ref{eq:1}) must be replaced by: 
\begin{equation}
    \frac{\mathrm{d}\rho}{\mathrm{d}t} = M^{(-1)}\mathrm{e}^{\mathrm{i}\omega_{\mathrm{mod}}t}\rho + M^{(0)}\rho + M^{(+1)}\mathrm{e}^{-\mathrm{i}\omega_{\mathrm{mod}}t}\rho - R\rho + N\ ,
\end{equation}
where $M^{(0)}$, $M^{(-1)}$, and $M^{(+1)}$ are $16\times 16$ matrices containing the terms in the Hamiltonian corresponding to the different frequency components of the field.

The solution to this equation can be approximated by performing a Floquet expansion of the density matrix, namely $\rho = \sum_{n} \rho^{(n)}\mathrm{e}^{-\mathrm{i}\omega_{\mathrm{mod}}t}$.   
The amplitudes $\rho^{(n)}$ of the different frequency components can then be obtained using the continuous fraction method \cite{wongInfluenceCoherentRaman2004}. By truncating these continuous fractions at orders $\pm 1$, one obtains the following system of equations:
\begin{equation}\label{eq:3}
\left\lbrace
\begin{aligned}
     \rho^{(0)} &= \bigl[R - M^{(0)} + M^{(+1)}(M^{(0)}-R-\mathrm{i}\omega_\mathrm{mod}\mathbb{I})^{-1} M^{(-1)} \\
     &\quad + M^{(-1)}(M^{(0)}-R+\mathrm{i}\omega_\mathrm{mod}\mathbb{I})^{-1}\bigr] N\ , \\
     \rho^{(-1)} &= -(M^{(0)}-R+\mathrm{i}\omega_\mathrm{mod}\mathbb{I} )^{-1} M^{(+1)}\rho^{(0)}\ , \\
     \rho^{(+1)} &= -(M^{(0)}-R+\mathrm{i}\omega_\mathrm{mod}\mathbb{I} )^{-1} M^{(-1)}\rho^{(0)}\ ,
\end{aligned}
\right.
\end{equation}

in which $\mathbb{I}$ is the $16\times 16$ identity matrix. After obtaining the matrices $\rho^{(0)}$, $\rho^{(-1)}$, and $\rho^{(+1)}$ from Equations (\ref{eq:3}), we can write the coherence between  states $\left| 1 \right>$ and $\left| 2 \right>$ in the rotating frame according to:
\begin{equation}\label{eq:4}
    \rho_{21} = \rho_{21}^{(-1)}\mathrm{e}^{\mathrm{i}\omega_\mathrm{mod}t} + \rho_{21}^{(0)} + \rho_{21}^{(+1)}\mathrm{e}^{-\mathrm{i}\omega_\mathrm{mod}t}\ .
\end{equation}

Equation (\ref{eq:4}) shows that the atoms radiate light at frequencies $\omega_{\mathrm{p}}$ and $\omega_{\mathrm{p}}\pm\omega_{\mathrm{mod}}$. This transfer of modulation from the coupling to the probe beam, which can be interpreted as a four-wave mixing phenomenon mediated by the nonlinearity of the atoms \cite{ducloyTheoryDegenerateFourwave1982}, adds sidebands to the probe beam. \textcolor{changecolor}{For example, we see that the  combination of the three incident frequencies $\omega_{\mathrm{c}}$, $\omega_{\mathrm{c}}\pm\omega_{\mathrm{mod}}$, and $\omega_{\mathrm{p}}$ creates the new side band frequency $\omega_{\mathrm{p}}\pm\omega_{\mathrm{mod}}$}. 

\textcolor{changecolor}{The appearance of these new sidebands induces a modulation of the intensity of the probe beam}, as depicted in Fig.\,\ref{fig:principle}(d). We can thus extract from the probe beam detection a novel signal known as the Relative Modulation Amplitude (R.M.A), defined as:
\begin{equation}
    \mathrm{R.M.A} = 2\left|\mathcal{E}_\mathrm{p}^{(0)}\left(\mathcal{E}_\mathrm{p}^{(-1)}\right)^* + \left(\mathcal{E}_\mathrm{p}^{(0)}\right)^*\mathcal{E}_\mathrm{p}^{(+1)}\right|\Big/\Big|\mathcal{E}_\mathrm{\mathrm{p}}^\mathrm{in}\Big|^2 \ ,\label{Eq09}
\end{equation}
where $\mathcal{E}_\mathrm{p}^{(0)}$, $\mathcal{E}_\mathrm{p}^{(-1)}$, and $\mathcal{E}_\mathrm{p}^{(+1)}$ are the complex amplitudes of the probe carrier and sidebands, respectively, and $\mathcal{E}_\mathrm{p}^\mathrm{in}$ is the complex amplitude of the incident probe field. It is worth noting that this signal, although obtained by demodulating the probe power at the phase modulation frequency of the coupling beam, differs significantly from what could be obtained using a standard lock-in detection method.

\begin{figure}[tbp]
\includegraphics[width = \columnwidth]{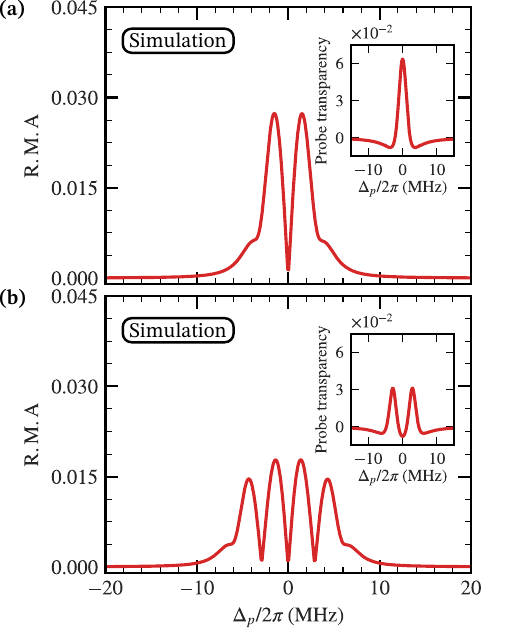}
\caption{\label{fig:Figure2} Simulated evolution of the Relative Modulation Amplitude (R.M.A.) of the probe intensity versus probe detuning for (a) $E_{\mathrm{RF}}=0$ and (b) $E_{\mathrm{RF}}=0.65\,\mathrm{V/m}$. The insets show the corresponding evolutions of the probe transparency in the Conventional Protocol. The transparency is defined as the variation in probe transmission induced by the presence of the coupling beam. The values of the simulation parameters are given in Table\,\ref{tab01}. For the Modulation Transfer Protocol, $\omega_\mathrm{mod}/2\pi=3$ MHz and $\beta = 0.25$.
}
\end{figure}

The simulated R.M.A signal of Eq.\,(\ref{Eq09}) is plotted in Fig.\,\ref{fig:Figure2} as a function of the probe detuning. The values of the parameters used to compute these spectra, as listed in Table\,\ref{tab01}, correspond to the experimental parameters that will be described below. Figure \ref{fig:Figure2}(a) shows the R.M.A spectrum in the absence of RF field. Contrary to the usual EIT spectrum shown in the inset, the R.M.A spectrum exhibits a very sharp destructive interference feature at $\Delta_{\mathrm{p}}=0$. This destructive behavior originates from the $\pi$ phase shift between the two side-bands of the phase modulated coupling beam. In the presence of an RF field resonant with the Rydberg transition, the signal becomes as shown in Fig.\,\ref{fig:Figure2}(b). In the following, we want to exploit the sharp resonance exhibited by these signals due to the destructive interference phenomena to enhance the bandwidth of the RF signal detection.

The modulation of the coupling field is based on two key parameters: i) the modulation frequency $\omega_\mathrm{mod}$, and ii) the modulation amplitude. This latter parameter is quantified by $\beta$ defined as:
\begin{equation}
\beta=\frac{\left|\mathcal{E}_\mathrm{c}^{(\pm1)}\right|^2}{2\left|\mathcal{E}_{\mathrm{c}}^{(\pm1)}\right|^2+\left|\mathcal{E}_\mathrm{c}^{(0)}\right|^2} .
\end{equation}

In the following section, we optimize these two parameters to maximize the RF detection bandwidth using the modulation transfer protocol.

\section{Experimental and theoretical optimization of the modulation transfer protocol}\label{sec:optimization}
Based on the theoretical approach described in the preceding section, the aim of the present section is to describe the experiments and simulations that allowed us to optimize the response of the modulation transfer protocol.
\subsection{Experimental Apparatus}
\begin{figure*}[t]
\includegraphics[width = \textwidth]{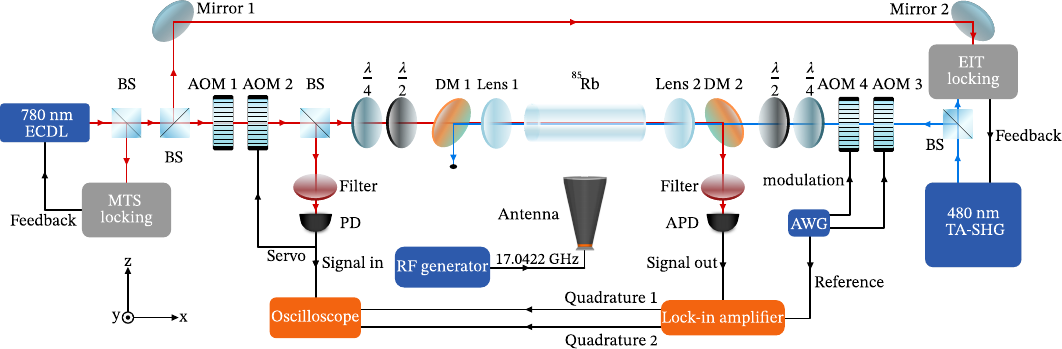}
\caption{\label{fig:Figure3}Simplified experimental setup. BS: beam splitter, $\frac{\lambda}{4}$: quarter-wave plate, $\frac{\lambda}{2}$: half-wave plate, PD: amplified photodetector, APD: avalanche photodiode, DM: dichroic mirror, AOM: acousto-optic modulator, AWG: arbitrary waveform generator. The amplified photodetector (PD) and the avalanche photodiode (APD) are used to monitor the probe laser power both before and after it passes through the vapor cell. In the conventional protocol, the coupling laser is modulated in amplitude using a square wave signal. In contrast, the modulation transfer protocol employs phase modulation of the coupling laser by a sinusoidal signal.}
\end{figure*}
Before discussing the R.M.A. optimization itself, let us first describe the experimental configuration, as represented in Fig.\,\ref {fig:Figure3}.

The probe and coupling lasers at 780 and 480\,nm, respectively, propagate in opposite directions through a quartz cell, which contains a natural mixture of Rubidium isotopes at room temperature. Within the cell, the probe (coupling) laser has a waist diameter of 300 (400)\,$\mathrm{\upmu m}$, leading to Rayleigh ranges longer than the 7.5~cm length of the vapor cell. The input laser powers are $0.4 \,\upmu\mathrm{W}$ for the probe laser and $46\,\mathrm{mW}$ for the coupling laser. A horn antenna (MVG QR18000) located at 57.5~cm from the vapor cell generates the RF field, which is sent to the cell  orthogonally to the laser beams. To shield the antenna and vapor cell from external fields, we use a custom-built small anechoic chamber made of aluminum sheets and lined with RF-absorbing panels. The probe, coupling, and RF fields are all linearly polarized along the Y-axis.

The frequency of the 780~nm probe laser (Toptica DL-Pro) is locked to the transition between levels $5^2S_{1/2}(F = 3)$ and $5^2P_{3/2}(F=4)$ of $^{85}$Rb using modulation transfer spectroscopy \cite{mccarronModulationTransferSpectroscopy2008, sunModulationTransferSpectroscopy2016, preuschoffOptimizationStrategiesModulation2018,limModulationTransferSpectroscopy2022} in an auxiliary cell. The 480 nm coupling laser (Toptica TA-SHG) is locked to the transition between the $5^2P_{3/2}(F=4)$ and $50^2D_{5/2}(F=3,4,5)$ levels using an EIT resonance in the presence of the probe laser \cite{abelLaserFrequencyStabilization2009}. 
The amplitude of the RF electric field $E_{\mathrm{RF}}$ at the position of the vapor cell is estimated after accounting for the losses, the gain of the antenna, and its distance from the vapor cell. The cell perturbation factor \cite{robinsonAtomicSpectraSixlevel2021, fanEffectVaporCellGeometry2015, rotunnoInvestigatingElectromagneticallyInduced2023}, which reduces the average RF field within the cell, is estimated to be approximately 0.54 according to experimental characterization using AT splitting. This RF signal couples level $50^2D_{5/2}(F=3,4,5)$ with level $51^2P_{3/2}(F=2,3,4)$.

To tune its frequency, the coupling beam passes through two single-pass AOMs inducing frequency shifts of opposite signs. The second AOM is also used to phase modulate \cite{liOpticalPhaseShifting2005} the coupling beam in the modulation transfer protocol. The phase modulation frequency $\omega_\mathrm{mod}$ and amplitude $\beta$ are programmed using an arbitrary waveform generator. \textcolor{changecolor}{This allows us to fine-tune $\beta$ and to only provide first-order sidebands to the modulation.} The value of $\beta$ is experimentally calibrated by measuring the laser spectrum with a Fabry-Perot interferometer. In the modulation transfer protocol, the probe beam intensity is demodulated at $\omega_\mathrm{mod}$, and the two quadratures provided by this demodulation are combined to measure the R.M.A defined in Eq.\,(\ref{Eq09}). In the conventional protocol, the coupling beam intensity is modulated by a square signal at 100~kHz with a 100\,\% modulation depth, and the probe beam intensity is demodulated with a lock-in detection. This directly provides the so-called probe beam transparency, which is the difference in probe beam transmission when the coupling beam is present and when it is absent. Similarly to the coupling beam, the probe beam frequency can also be scanned thanks to two acousto-optic modulators. In all cases, the probe beam is detected with a fast avalanche photodiode (APD) (Thorlabs APD410A/M) with a 10~MHz bandwidth and $1.2 \times 10^7 \ \mathrm{V/W}$ sensitivity at 780~nm. Another photodetector detects a fraction of the incident probe power on the cell to calibrate transmission measurements.

\begin{figure}[tbp]
\includegraphics[width = \columnwidth]{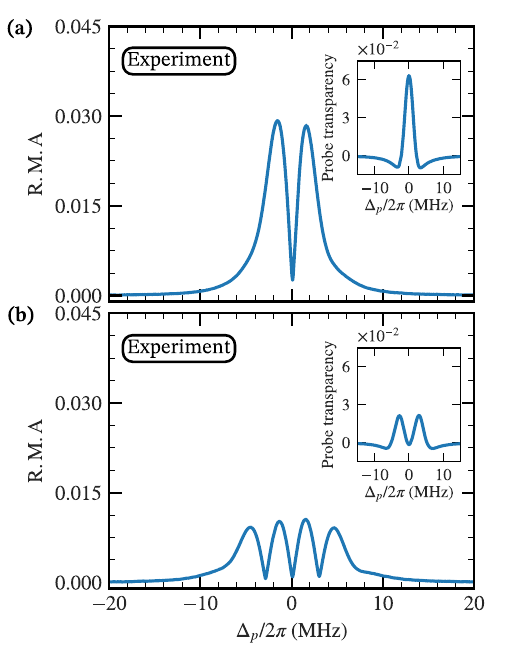}
\caption{\label{fig:exp_MTP} Experimental spectra corresponding to the simulations of Fig.\,\ref{fig:Figure2}.}
\end{figure}

Figure\ \ref{fig:exp_MTP} reproduces typical experimental spectra, obtained by scanning the probe laser frequency. Figures \ref{fig:exp_MTP}(a) and  \ref{fig:exp_MTP}(b) correspond to the R.M.A signal in the absence and in the presence of an RF field, respectively. The insets are the corresponding transparency spectra obtained in the conventional protocol, i.e., in the absence of phase modulation of the coupling beam and by measuring the DC transmission of the probe beam. The amplitudes and widths of both the R.M.A signal and the probe transparency of Fig.\,\ref{fig:exp_MTP}(a) are in very good agreement with the corresponding simulations of Fig.\,\ref{fig:Figure2}(a). In the presence of the RF field, the agreement between theory and experiments (see Fig.\,\,\ref{fig:Figure2}(b) and \,\ref{fig:exp_MTP}(b)) degrades slightly. In particular, the experimental spectra become slightly broader than the predicted one. We attribute this broadening to the fact that while the model relies on only four levels, the experimental level structure is much more complicated, and the different Zeeman sublevels of the different levels exhibit different AT splittings. Despite this small discrepancy, we still rely on our simple four-level model to optimize the modulation transfer protocol below.

\subsection{Optimization of the Phase Modulation of the Coupling Beam}

Now that we have checked the validity of our simulations, we can use them to optimize the response of the modulation transfer protocol. In particular, we have shown in Ref.\,\cite{trinhModulationTransferProtocol2024} that the modulation transfer protocol is particularly interesting to enhance the bandwidth of RF detection. This improvement comes from the sharp dip that can be observed around $\Delta_\mathrm{p}/2\pi = 0$ in Figs.\,\ref{fig:Figure2} and \ref{fig:exp_MTP}. This dip, which originates from destructive interference between the two sidebands of the coupling beam, creates a steep slope near the probe resonance $\Delta_\mathrm{p}/2\pi = 0\,\mathrm{MHz}$. When the RF field is detuned, one can use this steep slope at   $\Delta_\mathrm{p}/2\pi$ slightly different from 0 (typically $100\,\mathrm{kHz}$), to enhance the response of the R.M.A signal to the variations of $E_\mathrm{RF}$.

\begin{figure}[tbp]
\includegraphics[width = \columnwidth]{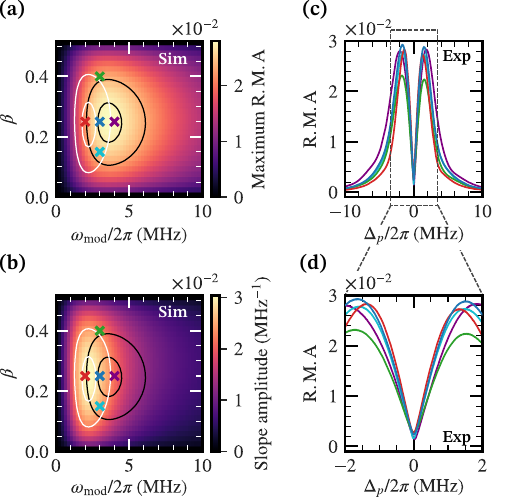}
\caption{\label{fig:optimization}Simulated  R.M.A signal (a) amplitude and (b) slope at $\Delta_{\mathrm{p}}=100\,\mathrm{kHz}$ versus $\beta$ and $\omega_\mathrm{mod}$. Black lines: iso-amplitudes (0.027 and 0.023). White lines: contours of constant signal slope (0.029 and 0.023 $\mathrm{MHz}^{-1}$). The colored crosses denote the parameters used to plot figures (c) and (d), which display experimental spectra of the MTP. (d) is a zoomed-in version of (c).}
\end{figure}

To optimize the system response in the modulation transfer protocol, we thus optimize the slope of the R.M.A at $\Delta_\mathrm{p}/2\pi=100\,\mathrm{kHz}$ and the amplitude of the R.M.A signal, by varying the values of $\omega_\mathrm{mod}$ and $\beta$. The corresponding results are summarized in Fig.\,\ref{fig:optimization}. Figure\,\ref{fig:optimization}(a) shows the simulated evolution of the R.M.A amplitude as a function of $\omega_\mathrm{mod}$ and $\beta$, as illustrated by the different experimental spectra shown in Fig.\,\ref{fig:optimization}(c). The maximum occurs for $\omega_\mathrm{mod}/2\pi = 3.5\,\mathrm{MHz}$ and $\beta = 0.25$, as shown by the black contour plots in Fig.\,\ref{fig:optimization}(a).  Similarly, the evolution of the R.M.A slope is shown in Fig.\,\ref{fig:optimization}(b), with a selection of associated measured spectra shown in Fig.\,\ref{fig:optimization}(c, d). This time, the maximum slope is obtained for $\omega_\mathrm{mod}/2\pi = 2\,\mathrm{MHz}$ and $\beta = 0.25$, as shown by the white contour plots in Fig.\,\ref{fig:optimization}(b).

As an acceptable trade-off between the two optima, we choose the values $\omega_\mathrm{mod}/2\pi = 3\,\mathrm{MHz}$ and $\beta = 0.25$. One notices that the optimized value of $\omega_\mathrm{mod}$ is of the order of $\gamma_{\mathrm{t}} + \gamma_{12}$, i.e., the order of magnitude of the width of the EIT peak, thus optimizing the overlap between the signals that must interfere destructively. Additionally, we note that the optimal value of $\beta$ equal to 0.25 means that half of the power of the coupling beam is contained in its two sidebands. Moreover, the simulations of Fig.\,\ref{fig:optimization} show that the modulation transfer protocol is quite robust to small variations of the parameters $\omega_\mathrm{mod}$ and $\beta$. As long as $\beta$ is between 0.15 and 0.4 and  $\omega_\mathrm{mod}/2\pi$  between 1~MHz and 6~MHz, the R.M.A signal remains comfortable.

\begin{figure*}[htp]
\includegraphics[width = \textwidth]{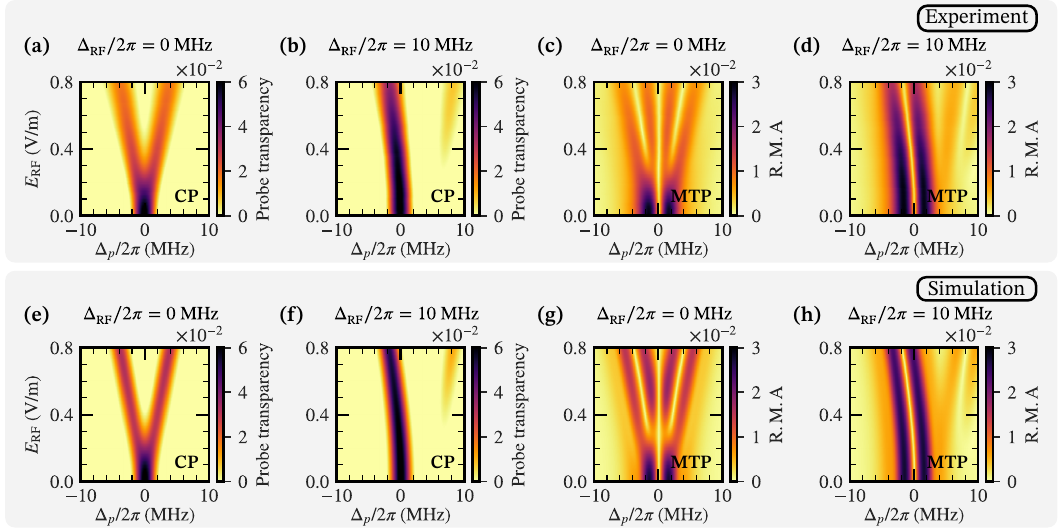}
\caption{Spectra evolutions versus (a, c, e, g) resonant or (b, d, f, h) detuned RF field amplitude. (a-d) Experiments. (e-h) Simulations. (a,~b,~e,~f)~Conventional Protocol. (c, d, g, h) Modulation Transfer Protocol.\label{fig:scans}}
\end{figure*}

Now that we have used the model of Section \ref{sec:theory} to optimize the parameters of the phase modulation of the coupling beam, we can check that the system indeed responds to a resonant or detuned RF field in the expected way. This is performed in Fig.\,\ref{fig:scans}, which compares simulated and experimental spectra for both the conventional and modulation transfer protocols and for both resonant and detuned RF fields. In every figure, the probe transparency (for the CP) or the probe R.M.A (for the MTP) are represented in color maps as a function of the probe detuning and of the amplitude of the RF field.

First of all, we observe that the simulations are in very good agreement with the experiments. Second, one can notice in Figs.\,\ref{fig:scans}(a) and \ref{fig:scans}(e) that the CP is very efficient to detect a resonant RF field. For example, by monitoring the probe transparency for $\Delta_{\mathrm{p}}=0$, one can see that the transparency undergoes a strong variation for field amplitudes as small as 0.2\,V/m. On the contrary, Figs.\,\ref{fig:scans}(b, f) confirm that the CP becomes quite inefficient as soon as the RF field is detuned by 10\,MHz from the transition between Rydberg levels. Conversely, Figs.\,\ref{fig:scans}(c, d, g, h) show that the exactly opposite conclusion holds for the MTP: at probe resonance ($\Delta_{\mathrm{p}}=0$), this protocol is insensitive to resonant RF fields (see Figs.\,\ref{fig:scans}(c, g)) but becomes extremely sensitive to small fields for detuned RF fields, for example when $\Delta_{\mathrm{RF}}/2\pi=10 \, \mathrm{MHz}$ (see Figs.\,\ref{fig:scans}(d,~~h)).

The above discussion, based on the results of Fig.\,\ref{fig:scans}, confirms that the optimized MTP permits to extend the sensitivity of the sensor to detuned RF fields. In the following section, we provide a quantitative comparison of the sensitivity of the two protocols and of the bandwidth improvement provided by the optimized MTP. \textcolor{changecolor}{We also demonstrate the MTP's ability to close the gap between two nearby RF transitions. }

\section{Expansion of RF bandwidth}\label{section:bandwidth}

This section explores the expansion of the RF bandwidth allowed by the MTP optimization performed in Section III.B. We will first examine the bandwidth expansion around a single transition. Next, we will compare the sensitivities of the two protocols. Finally, we will assess whether the bandwidth improvement is sufficient to bridge the frequency gap between two Rydberg transitions.

\textcolor{changecolor}{\subsection{Expansion of RF Bandwidth around a Single Transition}}

In order to examine the performance of the MTP in comparison to the CP, we focus on weak and detuned RF fields. In particular, we try to see whether the MTP permits to expand the RF bandwidth. The sensitivity $S$ \cite{schlossbergerRydbergStatesAlkali2024}, expressed in $\mathrm{V.m^{-1}.{Hz}^{-1/2}}$ is defined as    
\begin{equation}
    S = \frac{V_0}{\left(\frac{\partial V}{\partial E_\mathrm{RF}}\right)\sqrt{\mathrm{RBW}}}\ ,\label{eq11}
\end{equation}
where $\frac{\partial V}{\partial E_\mathrm{RF}}$ is the dependence of the measured voltage on the RF signal amplitude at $E_{\mathrm{RF}}=0$,  $V_0$  is the RMS signal noise in the detection bandwidth RBW. As discussed below, in our experimental conditions, this noise is of the same order of magnitude for the CP and the MTP.

\begin{figure}[tbp]
\includegraphics[width = \columnwidth]{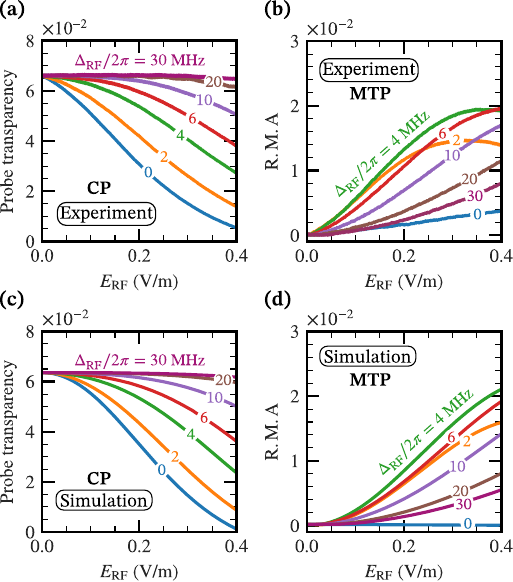}
\caption{\label{fig:transparency_vs_ERF}(a, b) Measured and (c, d) simulated sensor responses for various detunings of the RF field ($\Delta_\mathrm{RF}/2\pi = 0, 2, 4, 6, 10, 20, \mathrm{and} \ 30$ MHz). (a, c) Conventional Protocol with $\Delta_\mathrm{p}/2\pi = 0$ kHz. (b, d) Modulation Transfer Protocol with $\Delta_\mathrm{p}/2\pi = 100$ kHz. }
\end{figure}

The slope $\partial V/\partial E_\mathrm{RF}$ is determined by measuring the probe output signal with an avalanche photodiode at $\Delta_\mathrm{p} = 0$ in the conventional protocol and at $\Delta_\mathrm{p}/2\pi = 100$ kHz in the modulation transfer protocol. The output voltage is monitored while scanning $E_\mathrm{RF}$.
Examples of such signal responses to various detunings of the RF field are presented in Figs.\ref{fig:transparency_vs_ERF}(a, b) for the CP and the MTP protocols, together with the corresponding simulations in Figs.\ref{fig:transparency_vs_ERF}(c, d). The simulations and the experiment agree well, while the small remaining discrepancies are, as mentioned above, attributed to the hyperfine structure that is overlooked in the simulations.

As anticipated in Fig.\,\ref{fig:scans}, the change in probe transparency with $E_{\mathrm{RF}}$ is maximum at RF resonance for the CP. The slope of the CP response curve almost reaches zero for $\Delta_\mathrm{RF}/2\pi= 30$ MHz, resulting in a poor sensitivity. On the contrary, for the MTP, the signal at resonance shows a minimal response to the RF field. However, the slope gradually increases until $\Delta_\mathrm{RF}/2\pi= 4$ MHz before decreasing but remaining significant at $\Delta_\mathrm{RF}/2\pi= 30$ MHz.

\begin{figure}[tbp]
\includegraphics[width = \columnwidth]{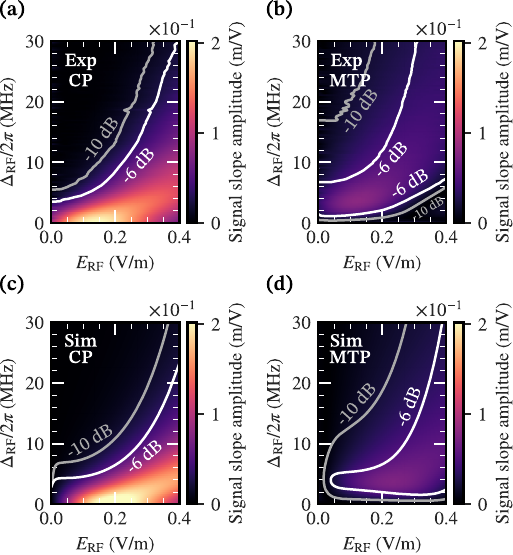}
\caption{\label{fig:slopes}(a, b) Experimental and (c, d) theoretical slopes of the response of the system to variations of the RF field amplitude, plotted versus $E_{\mathrm{RF}}$ and $\Delta_\mathrm{RF}/2\pi$. (a, c) Conventional Protocol with $\Delta_{\mathrm{p}} = 0$ kHz. (b, d) Modulation Transfer Protocol with $\Delta_{\mathrm{p}}/2\pi = 100\,\mathrm{kHz}$. The white and grey lines denote the RF bandwidth at -6 dB  and -10 dB, respectively, compared to the CP response at $\Delta_\mathrm{RF} = 0$.}
\end{figure}

To better appreciate the measurement bandwidth of the two protocols, we performed 101 measurements for $\Delta_\mathrm{RF}/2\pi$ ranging from 0 to 30\,MHz, and interpolated the results with polynomials. This leads to the response slopes represented as color maps in Fig.\,\ref{fig:slopes}.

In the conventional protocol, the strongest slope, and thus best sensitivity to the RF field is measured for $\Delta_\mathrm{RF}/2\pi = 0$ MHz and for values of $E_\mathrm{RF}$ between 0.05 V/m and 0.25 V/m. Based on this reference value, the white and grey contour lines in Fig.\,\ref{fig:slopes} correspond to decreases in the slope of 6 and 10 dB, respectively. These plots show that for the CP, the maximum sensitivity for detuned RF fields is obtained for higher values of $E_{\mathrm{RF}}$. Besides, in the MTP, the best sensitivity is achieved between $\Delta_\mathrm{RF}/2\pi = 2$ and 6 MHz. In this domain of detunings and above, the MTP outperforms the CP.

Figure\,\ref{fig:slopes} shows how the MTP allows for improving the bandwidth of the sensor. For vanishing RF field amplitudes, the RF bandwidth at -6 dB is $\Delta_\mathrm{RF}/2\pi = 3.5$ MHz in the conventional protocol, while it is $\Delta_\mathrm{RF}/2\pi = 6.5$ MHz in the modulation transfer protocol. Similarly, the RF bandwidth at -10 dB (indicating a loss of $90\%$ of sensitivity) is reached at $\Delta_\mathrm{RF}/2\pi = 5.5$ MHz in the conventional protocol and at $\Delta_\mathrm{RF}/2\pi = 17$ MHz in the modulation transfer protocol. This is an improvement of the RF bandwidth by 11.5 MHz. This improvement is slightly larger in the experiment than expected by the simulations.

\textcolor{changecolor}{\subsection{Sensitivity Comparison between the Two Protocols}}

Finally, we wish to determine in which range of parameters and to which extent the modulation transfer protocol outperforms the conventional protocol.

\begin{figure}[tbp]
\includegraphics[width = \columnwidth]{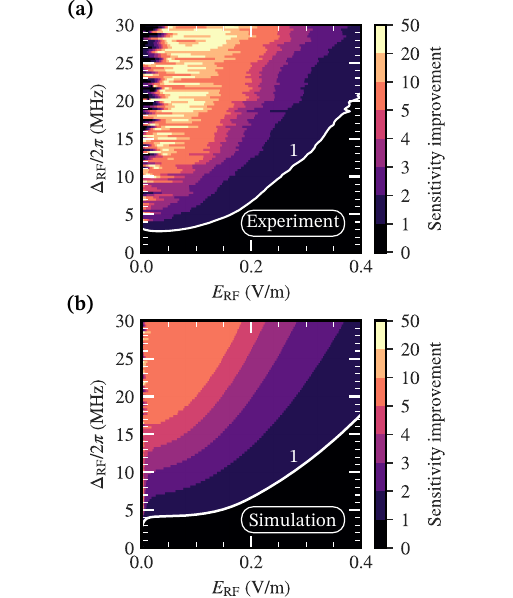}
\caption{\label{fig:slopes_ratio}Sensitivity improvement of the MTP versus CP sensitivity between the MTP and the CP for various amplitudes of the RF field and various RF detunings $\Delta_\mathrm{RF}$. (a) Experiment. (b) Simulation. The white lines indicate where the sensitivity levels for both protocols are equal.}
\end{figure}

To perform this comparison, since the noise associated with the two protocols is similar, we calculate the ratio between the slopes of the system responses in the two protocols. This ratio is shown in color maps in Fig.\ref{fig:slopes_ratio}, both experimentally and theoretically. The white lines indicate where the sensitivity levels for both protocols are equal, i.e., when the sensitivity improvement factor equals 1. Above that line, the modulation transfer protocol outperforms the conventional one. In the experiment (see Fig.\ref{fig:slopes_ratio}(a)), the modulation transfer protocol exhibits better performance as soon as $\Delta_\mathrm{RF}/2\pi = 3$ MHz for vanishing RF field amplitudes. Starting from $\Delta_\mathrm{RF}/2\pi = 4$ MHz, the improvement in sensitivity exceeds 2. This limit closely follows the RF bandwidth at -6 dB. For even greater RF detunings, this improvement goes beyond a factor of 20. As discussed previously, the simulations of Fig.\,\ref{fig:slopes_ratio}(b) closely replicate the experiment.

\begin{table}[ht!]
\caption{Sensitivities of the RF field amplitude measurement for the two measurement protocols and for several values of the RF detuning and $E_\mathrm{RF} = 0.05\, \mathrm{V/m}$. The sensitivities are given in $\mathrm{\upmu V.cm^ {-1}.Hz^{-1/2}}$. The first (second) part of the table is based on a noise value $V_0$ measured experimentally (deduced from the shot noise limit). }
\label{tab02}
\begin{ruledtabular}
\begin{tabular}{ccc}
Sensitivity Based on &  Conventional   &  Modulation Transfer  \\
($\mathrm{\upmu V.cm^ {-1}.Hz^{-1/2}}$) &   Protocol  &  Protocol  \\
\hline
\hline
Measured Noise &     &    \\
\hline
$\Delta_\mathrm{RF}/2\pi= 0 \,\mathrm{MHz}$ & 1.0 & 21.2\\ 
$\Delta_\mathrm{RF}/2\pi= 5 \,\mathrm{MHz}$& 7.4 & 1.3\\
$\Delta_\mathrm{RF}/2\pi= 10 \,\mathrm{MHz}$ & 36.0 & 2.6\\ 
$\Delta_\mathrm{RF}/2\pi= 20 \,\mathrm{MHz}$& 350.6 & 5.3\\
$\Delta_\mathrm{RF}/2\pi= 30 \,\mathrm{MHz}$ & 529.2 & 8.1\\ 
\hline
\hline
Shot Noise \\                                                                   
\hline
$\Delta_\mathrm{RF}/2\pi= 0 \,\mathrm{MHz}$ & 0.1 & 3.9\\ 
$\Delta_\mathrm{RF}/2\pi= 5 \,\mathrm{MHz}$& 0.6 & 0.2\\
$\Delta_\mathrm{RF}/2\pi= 10 \,\mathrm{MHz}$ & 2.8 & 0.5\\ 
$\Delta_\mathrm{RF}/2\pi= 20 \,\mathrm{MHz}$& 26.7 & 1.0\\
$\Delta_\mathrm{RF}/2\pi= 30 \,\mathrm{MHz}$ & 40.3 & 1.5\\ 
\end{tabular}
\end{ruledtabular}
\end{table}

We finally give in this last paragraph a comparison between the sensitivities, defined by Eq.\,(\ref{eq11}), of the two protocols. Table \ref{tab02} presents the sensitivities of the two protocols for various values of the RF detuning. For each case, we give two values of the sensitivity: one is obtained by taking the shot noise limit for $V_0$ in Eq.\,(\ref{eq11}), while the other is based on measurements of the noise $V_0$ in our experiment. In both cases, the slope $\partial V/\partial E_\mathrm{RF}$ is measured. We notice that the MTP outperforms the CP as soon as $\Delta_\mathrm{RF}/2\pi > 3\,\mathrm{MHz}$ for weak fields. Moreover, the improvement of the sensitivity for the MTP with respect to the CP is even better when we consider the measured noise than when we consider the shot noise limit. This is due to the fact that the MTP corresponds to a measurement at 3\,MHz, where the detected noise is much lower than at the lower frequency (100\,kHz) where the CP is implemented. Moreover, we observe that these sensitivity measurements compare very favorably with those in the existing literature \cite{fanAtomBasedRF2015,kumarAtomBasedSensingWeak2017} within the same range of RF frequencies. 

\textcolor{changecolor}{\subsection{Closing a Frequency Gap between Two Rydberg Transitions}} 

\textcolor{changecolor}{Now that we know that the MTP expands the RF bandwidth around a single transition, let us examine whether it can bridge the gap between two nearby Rydberg transitions.}

\begin{figure}[htp]
    \centering
    \includegraphics[width=\columnwidth]{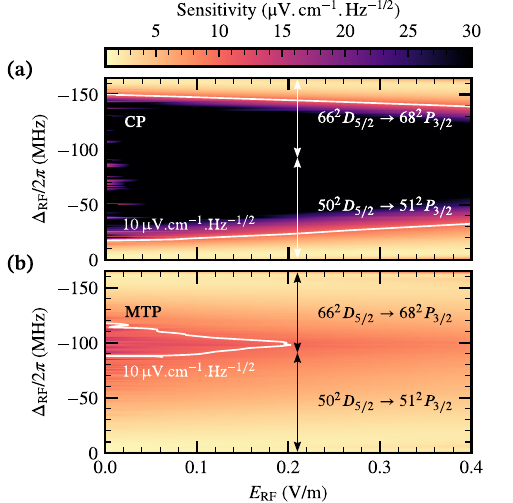}
    \caption{\textcolor{changecolor}{Evolution of the measured sensitivity to the RF field amplitude for various detunings between two nearby RF transitions. (a)~Conventional Protocol. (b)~Modulation~Transfer~Protocol. The white lines indicate an iso-sensitivity of $10 \, \upmu \mathrm{V.cm^{-1}.Hz^{-1/2}}$.}  }
    \label{fig:gap_closing}
\end{figure}

\textcolor{changecolor}{We selected a nearby transition, which couples $66^2D_{5/2}$ and $68^2P_{3/2}$ states (transition II), and is approximately 166\,MHz away from the previously used RF transition. This frequency gap is too large for the CP to maintain a good sensitivity between the two transitions, as shown by Fig.\,\ref{fig:gap_closing}(a). This figure shows, as a color map, the evolution of the measured sensitivity as a function of the detuning from the $50^2D_{5/2}\rightarrow 51^2P_{3/2}$ (transition I) transition (vertical axis) and of the RF field amplitude (horizontal axis). The two vertical white arrowed lines show the regions in which we tune the coupling laser close to the $50^2D_{5/2}$ level or to the $66^2D_{5/2}$ level: the detuning varies from $\Delta_\mathrm{RF}/2\pi=0$ to $\Delta_\mathrm{RF}/2\pi= - 100 \,\mathrm{MHz}$ for the first transition and from $\Delta_\mathrm{RF}/2\pi=0$ to $\Delta_\mathrm{RF}/2\pi=+66 \,\mathrm{MHz}$ for the second one. The dark region between the two transitions is a signature of the fact that the sensitivity becomes very poor as soon as the RF field is detuned by a few MHz from resonance. The white contour lines correspond to the parameter values for which the sensitivity is equal to $ S = 10 \, \upmu \mathrm{V.cm^{-1}.Hz^{-1/2}}$.  A good sensitivity (around $ 1 \, \upmu \mathrm{V.cm^{-1}.Hz^{-1/2}}$) is obtained only at 0 detuning and at $-166$\,MHz, corresponding to the resonances of the two transitions. Notice that we carefully chose two transitions with similar electric dipole moments. However, in order to compensate for the 30\% drop in dipole moment of the coupling transition, we increased the coupling laser power by a factor of two when using transition II.} \textcolor{changecolor}{One should notice that since we are using two slightly different RF frequencies, the RF pattern inside the cell is modified. We adjusted the position of the laser beams inside the cell to have a close cell perturbation factor. It was previously 0.54 and is now 0.59 for transition I. Also, the RF field is more homogeneous, leading to less peak broadening as the RF field amplitude increases.  This results in better sensitivities than in Table \ref{tab02}, from $36 \, \upmu \mathrm{V.cm^{-1}.Hz^{-1/2}}$ down to $4 \, \upmu \mathrm{V.cm^{-1}.Hz^{-1/2}}$ at $\Delta_\mathrm{RF}/2\pi= 10 \,\mathrm{MHz}$ with the CP and from $2.6 \, \upmu \mathrm{V.cm^{-1}.Hz^{-1/2}}$ down to $1.8 \, \upmu \mathrm{V.cm^{-1}.Hz^{-1/2}}$ with the MTP. For the new transition, the cell perturbation factor is 0.64.}


\textcolor{changecolor}{For comparison, Fig.\,\ref{fig:gap_closing}(b) reproduces the same measurements performed using the MTP. One can see that the gap between the two transitions has now completely disappeared. Although the sensitivity of the MTP at resonance is slightly decreased to about $30 \, \upmu \mathrm{V.cm^{-1}.Hz^{-1/2}}$, it becomes much better than this (down to $1 \, \upmu \mathrm{V.cm^{-1}.Hz^{-1/2}}$) for all other values of the detuning. This illustrates the interest of the MTP to mitigate the detrimental discrete nature of Rydberg resonances.}

\section{Conclusion}

In this article, we have developed a theoretical model for the quantum Rydberg receivers based either on the measurement of the DC power of the transmitted probe field (so-called conventional protocol) or on the measurement of the transfer of modulation from the coupling beam to the probe beam in the atoms (so-called modulation transfer protocol). By comparison with an experiment using $^{85}$Rb atoms, we have seen that this model accurately reproduces the experimental results. This model was used to optimize the two key parameters for the coupling beam phase modulation in the modulation transfer protocol, namely the modulation frequency $\omega_\mathrm{mod}$ and the modulation amplitude $\beta$. The theoretical and experimental results revealed that, under the experimental conditions used here, the optimal combination is $\omega_\mathrm{mod}/2\pi = 3$ MHz and $\beta = 0.25$.

An analysis of the performances of the two protocols, both theoretical and experimental, has shown that the conventional protocol is the most efficient one when the RF field is resonant with the Rydberg transition. However, its sensitivity decreases rapidly when the RF field detuning increases. In contrast, the optimized modulation transfer protocol demonstrates superior sensitivity starting from a detuning of $\Delta_\mathrm{RF}/2\pi = 3$ MHz and beyond. The improvement in sensitivity provided by this latter protocol can exceed a factor of 20 at large detunings. Furthermore, this protocol allows for an increase in RF bandwidth at -10 dB by 11.5~MHz, resulting in a bandwidth of 17~MHz.

The two protocols are thus complementary: the conventional protocol is recommended for RF field detunings smaller than 3 MHz, while the modulation protocol should be used for larger detunings. Since both protocols use the same apparatus, switching between them is straightforward and can be accomplished by modifying the modulation applied to the coupling beam AOM. One can also operate both protocols simultaneously within the same vapor cell. The optimized modulation parameters presented in this paper only apply to Rubidium 85; they should be tailored to any other Alkali atom and adapted when the number of probed atoms is modified. \textcolor{changecolor}{We demonstrated that the modulation transfer protocol enables closing RF frequency gaps of at least 166~MHz, which is promising for developing a frequency-continuous sensor. The results presented in this article could be improved by heating the cell, using optical repumping, or increasing the atoms' transit time.} However, it is important to note that this technique does not achieve RF bandwidths or sensitivities comparable to alternative techniques based on the use of an RF local oscillator \cite{andersonContinuousfrequencyMeasurementsHighintensity2017,huContinuouslyTunableRadio2022,yaoSensitivityEnhancementFardetuned2022,liRydbergAtombasedAM2022, simonsContinuousRadiofrequencyElectricfield2021}. \textcolor{changecolor}{It is also not sensitive to the phase of the RF field.} Nonetheless, the protocol presented in this article is an all-optical method that does not require any external RF antenna, which would undermine the goal of developing a dielectric sensor.

Future developments of the modulation transfer protocol may involve exploring different modulation patterns or using the quadratures separately to investigate various nonlinear responses. Additionally, combining the modulation transfer protocol with other methods that allow for measuring the phase or polarization of the RF field also deserves investigation. 
\\
\begin{acknowledgments}
 We acknowledge the significant contributions of Joseph Delpy and Nadia Belabas, whose insightful discussions have profoundly impacted our work. Additionally, we recognize Sébastien Rousselot for his vital technical support.
This work was funded by the French Defense Innovation
Agency, Quantum Saclay, Agence Nationale de la Recherche (ANR) for France 2030 under Grant Agreement No.~ANR-23-PETQ-0004 and the European Defense Fund (EDF) under Grant Agreement No.~EDF-2021-DIS-RDIS-ADEQUADE (No.~101103417).
This study was funded by the European Union. Views and
opinions expressed are, however, those of the author(s) only and do
not necessarily reflect those of the European Union or the European
Commission. Neither the European Union nor the granting
authority can be held responsible for them.
\end{acknowledgments}

\bibliography{biblio.bib}

\end{document}